\documentclass[12pt, draftclsnofoot, onecolumn]{IEEEtran}

\usepackage{amsfonts, amsthm}
\usepackage{amsmath, color}
\usepackage{graphicx}
\usepackage{cite}
\usepackage[normalem]{ulem}
\usepackage{tikz}
\usepackage{epstopdf}
\usepackage{psfrag}
\usepackage{etoolbox}
\usepackage[process=auto]{pstool}
\usetikzlibrary{arrows,automata}
\usepackage{graphicx}
\usepackage{amssymb}
\usepackage{siunitx}

\newtheorem{remark}{Remark}

\allowdisplaybreaks
\title{Diffusive Molecular Communication in Biological Cylindrical Environment}
\author{Mohammad~Zoofaghari, Hamidreza~Arjmandi\\
Electrical Engineering Department, Yazd University
}

\begin{document}
\maketitle
%
%


\begin{abstract}
Diffusive molecular communication (DMC) is one of the most promising approaches for realizing nano-scale communications in biological environments for healthcare applications. In this paper, a DMC system in biological cylindrical environment is considered, inspired by blood vessel structures in the body. The internal surface of the cylinder boundary is assumed to be covered by the biological receptors which may irreversibly react with hitting molecules. Also, information molecules diffusing in the fluid medium are subject to a degradation reaction and flow. The concentration Green's function of diffusion in this environment is analytically derived which takes into account asymmetry in all radial, axial and azimuthal coordinates. Employing obtained Green's function, information channel between transmitter and transparent receiver of DMC is characterized. To evaluate the DMC system in the biological cylinder, a simple on-off keying modulation scheme is adopted and corresponding error probability is derived. Particle based simulation results confirm the proposed analysis. Also, the effect of different system parameters on the concentration Green's function are examined. Our results reveal that the degradation reaction and the boundary covered by biological receptors may be utilized to mitigate intersymbol interference and outperform corresponding error probability. 

\end{abstract}


\section{Introduction}

Diffusive molecular communication (DMC) is a promising approach for realizing nano-scale communications \cite{Akyldiz2011}. In DMC, information is encoded in the concentration, type, and/or release time of  molecules. In particular, a transmitter nanomachine releases information molecules into the environment. The released molecules move randomly via Brownian motion and may be observed at the receiver \cite{Pierobon2011}. DMC is an attractive communication mechanism for healthcare applications due to the possibility of bio-compatibility \cite{Nakano12}-\cite{Farsad16}. In biological environment, various phenomena influence the performance of DMC system which should be taken into account, e.g. boundaries covered by biological receptors, chemical reactions within the fluid involving information molecules, and the flow of the fluid medium.

Diffusion in an ideal unbounded environment has been extensively studied in MC literature\cite{Kuran10}-\cite{Assaf}. 
However, the unbounded environment which has a limited range of communication is not a realistic assumption for MC applications, particularly in in-vivo environment.
Diffusion in bounded spherical environment has also been considered in some works. The authors in \cite{Alzubi18} consider a DMC system in bounded spherical environment  where the transmitter is a point source, the receiver is located at the center of the sphere and covered by ligand receptors, and the sphere boundary is  pure absorbing. In \cite{Dinc182}, a spherical environment with reflective boundary is considered for DMC with a spherical absorbing receiver and a point source transmitter.

As a more compatible model for the blood vessel structures in the body and microfluidic channels, a bounded cylindrical environment can be employed. In particular, modeling the complicated shape of blood vessels as cylinders is well accepted \cite{Four}.
Investigation of MC in the cylindrical environment adopting various assumptions has been considered in the literature.
In \cite{Farsad12} a MC system in confined space of a microfluidic chip with constant flow is considered to compare different propagation schemes including free diffusion and active transport in terms of the achievable rates. The authors characterize communication channel for Brownian motion inside the confined environment with elastic walls by using particle based simulation. In \cite{Kuran13}, a DMC model is proposed in which a tunnel-like environment without flow composed of destroyer molecules exist between the transmitter and receiver to decrease the variance of the hitting times and obtain better signal shape. The hitting times and probabilities are obtained based on simulation results.
The response to a pulse of carriers released by a mobile transmitter, measured on a number of receivers located along the vessel wall based on simulation results is studied in \cite{Feli13}. In \cite{Feli14}, the authors propose a mathematical model for MC between platelets and endothelial cells via CD40 signaling based on a two-dimensional Markov chain. This model along with typical propagation models of blood vessels, has been incorporated into a simulation platform. In \cite{Turan18}, a tunnel-like environment without flow for DMC is considered where the receiver partially covers the cross-section of the tunnel, and the tunnel boundaries reflect the molecules upon contact. The distribution of hitting locations is obtained based on simulation results. 
The authors in \cite{Felicetti13} present a software platform for simulation of DMC within the blood vessels with drift. Also, in \cite{Unterweger}, an experimental testbed for molecular communication in duct flow has been presented using magnetic nanoparticles. The authors in \cite{Schefer} propose a semi-analytical model for a DMC channel in the cylindrical environment in the presence of a magnetic force and flow.

The authors in \cite{Wayan17} consider the diffusion environment as a straight cylindrical duct with reflective boundary which is filled with a fluid with non-uniform flow. Assuming a transmitter point source, the channel impulse responses for two simplifying flow regimes called dispersion and flow-dominant are obtained.
The authors in \cite{Dinc18} consider a 3-D microfluidic channel environment in the presence of flow where the boundaries are reflective. By employing the symmetry in azimuthal coordinates, the authors derive channel impulse response in radial and axial coordinates. 
The channel impulse response analysis provided in both works \cite{Wayan17}-\cite{Dinc18} are not able to consider the asymmetry of diffusion in azimuthal coordinate which may be unavoidable depending on the locations of transmitter and receiver. Moreover, a simple reflective boundary is assumed in both works. This assumption may not be realistic in in-vivo environments, where boundaries may be covered by biological receptors leading to partially absorption of molecules. For instance, the inner layer of blood vessels is composed of endothelium cells whose surface contains various types of receptors \cite{Cliff}.

As another important mechanism in in-vivo environment, the effect of degradation reactions has not been analytically investigated for DMC in the cylindrical environment, in the previous works. Interestingly, the degradation reaction may be utilized to overcome intersymbol interference (ISI) of the diffusion channel \cite{Feli13},\cite{Adam14} which is inspired from living organisms. For instance, Acetylcholinesterase molecules destroy the messenger Acetylcholine molecules in the channel between nerve cell and muscle cell in neuromuscular junction to clean the channel for the next signal transmission \cite{Nelson}.

In this paper, we consider a point-to-point DMC system(a DMC with one transmitter and one receiver) in a \textit{biological cylindrical environment}. A cylinder with partially absorbing boundary is assumed which is fully covered by the biological receptors. Information molecule (ligand) hitting to the boundary may react and bind to the receptor and produce a ligand-receptor complex. Simply, an irreversible ligand-receptor reaction is considered which makes our analysis analytically tractable. Further, a degradation reaction is assumed within the environment in which the diffusive information molecule may be transformed into another type with a probability depending on the reaction constant. Moreover, a uniform flow \cite{Farsad12} in the cylinder is assumed. Although, flow with constant velocity is not a realistic assumption in cylindrical environment, it makes possible to obtain analytic solution for corresponding diffusion equations which provides insightful ideas about the effect of different system parameters.
Concentration Green's function (CGF) of diffusion in this environment is analytically derived in terms of a convergent infinite series which takes into account asymmetry in all radial, axial, and azimuthal coordinates.

A point-to-point DMC system is assumed in this biological cylindrical environment. Employing obtained CGF, the average received signal at the observer receiver is obtained given an arbitrary transmitter geometry (not necessarily point source) with arbitrary transmitted modulated signal (not necessarily impulsive release signal). Further, the noise at the receiver is analyzed and information channel between the transmitter and receiver is characterized, accordingly. To evaluate the proposed DMC system, a simple on-off keying modulation scheme is adopted and corresponding error probability is derived. 
Noteworthy, obtained CGF for considered simplified model of the blood vessel provides insightful ideas for prediction of the concentration profile of drug in the blood vessels for healthcare applications.
 \color{black}

Our particle based simulation (PBS) results confirm the proposed analysis. Also, the effect of different system parameters on the CGF are examined and discussed. 
It is observed that the degradation reaction and partially absorbing boundary may be utilized to mitigate ISI and outperform corresponding error probability.
Moreover, our results reveal that the analysis proposed for the CGF with constant velocity flow well approximates CGF obtained from PBS with more realistic Poiseuille flow model, for enough small velocity values.

The paper is organized as follows. The system model is presented in Section II. The CGF of diffusion in the biological cylindrical environment is obtained in Section III. In Section IV, the information channel between the transmitter and receiver is characterized and the error probability of DMC with a simple on-off keying modulation over this channel is derived. The simulation and numerical results are presented in Section V. Finally, the paper is concluded in Section VI.




\section{System Model} \label{system model} \label{section2}
A point-to-point DMC system is considered within a biological cylindrical environment. Cylindrical coordinate system is employed to describe the environment where $(\rho,z,\varphi)$ denote radial, axial, and azimuthal coordinates, respectively. An infinite-height cylinder is assumed where the geometric location of points over the cylinder boundary is described as all points $(\rho,z,\varphi)$ where
\begin{equation}
	\rho=\rho_c,\;\;\; -\infty<z<+\infty, \;\;\; 0\leq \varphi<2\pi.
\end{equation}
The cylinder is filled with a fluid medium with the diffusion coefficient $D$ ($\si{m^2.s^{-1}}$) for information molecules $\mathrm{A}$. Diffusion coefficient is assumed uniform in all directions. \color{black} It is assumed the information molecules released in the environment may be degraded with a probability and transform to another molecule type under the following first order degradation reaction
 \begin{equation}\label{deg1}
	\mathrm{A} \overset{k_{d}}{\,\,\to} \mathrm{\hat{A}},
\end{equation}
where $k_d$ is the degradation reaction constant in \si{s^{-1}} and molecule $\mathrm{\hat{A}}$ is not recognized by the receiver.
The cylinder boundary is assumed to be fully covered by infinitely many biological receptors where hitting information molecule (ligand) may react and bind to the receptor (R) and produce a ligand-receptor complex (AR). We assume a simple irreversible reaction for the receptors on the boundary as
 \begin{equation} \label{deg2}
	\mathrm{A}+\mathrm{R} \overset{k_{f}}{\to} \mathrm{AR},
\end{equation}
where $k_{f}$ is forward reaction constant in \si{m.s^{-1}}.
It is obvious that the boundary is pure reflective and absorbing, for $k_{f}=0$ and $k_{f}=\infty$, respectively. We note that the effect of receptor occupancy is neglected and the formations of the individual ligand-receptor
complexes are assumed independent of each other. Consequently, multiple information molecules can react on the boundary at the same time and at the same location.

A flow with constant velocity $v$ \si{m.s^{-1}} \cite{Farsad12} in axial direction is considered inside the cylinder, i.e., the velocity field is given by $\bar{v}(\bar r)=v \hat a_z $ \si{m.s^{-1}} where $\hat a_z$ is unit vector in axial direction.\color{black} 

\begin{figure}
\center
\includegraphics[width=17 cm,height=10 cm]{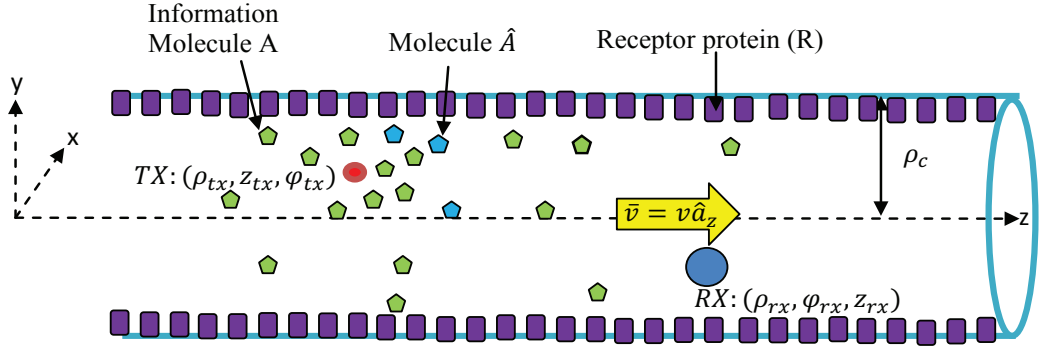}
	\setlength{\abovecaptionskip}{-0.5 cm}
 \caption{Schematic illustration of considered DMC system in biological cylindrical environment.}
\label{Fig0}
\end{figure}

The transmitter is assumed to be a point source located inside the cylinder at an arbitrary point $\bar{r}_{\rm tx}=(\rho_{\rm tx},z_{\rm tx},\phi_{\rm tx})$ where $0\leq \rho_{\rm tx} \leq \rho_c$. The transmitter uses information molecule of type A.
Also, a transparent spherical receiver with radius $R_{\rm rx}$ and center at $\bar{r}_{\rm rx}=(\rho_{\rm rx},z_{\rm rx},\varphi_{\rm rx})$ is considered that does not affect the Brownian motion of molecules.
%
A schematic illustration of the system model is represented in Fig. \ref{Fig0}

A time-slotted communication scenario with the time slot duration of $T$ seconds (s) is considered. The receiver and transmitter are assumed perfectly synchronized. In fact, it is possible to synchronize a transmitter with a receiver in molecular communication at the cost of an overhead \cite{syncArj}\color{black}. The transmitter controls (modulates) the average release rate of molecules into the environment according to the input symbol. The modulated signal corresponding with a symbol $i$ is denoted by $s^i(t)$. The released molecules diffuse in the environment following a Brownian motion and their movements are assumed to be independent of each other. The receiver counts the number of molecules falling inside its volume at a sampling time $t_s$ to decide about the intended transmitted symbol.

To analyze the presented DMC system, we first obtain CGF of diffusion channel. The information channel between the transmitter and receiver in this environment is characterized, correspondingly. Then, the stochastic of the received signal at the receiver is investigated. In order to evaluate the performance of the system in terms of error probability, a simple on-off keying modulation is adopted. Bits $1$ and $0$ are represented by the instantaneous release of $N$ molecules (on average) and no molecule, respectively.


\section{CGF of diffusion in biological cylindrical environment}
In this section, we derive concentration Green's function for diffusion inside the cylinder described above.

To this end, we assume the point source transmitter located at $\bar{r}_{\rm tx}=(\rho_{\rm tx},z_{\rm tx},\varphi_{\rm tx})$ with instantaneous molecule release rate $\delta(t-t_0)$ molecule \si{\rm mol.s^{-1}}, where $\delta(.)$ is Dirac delta function and $t_0$ is constant representing the release instant. Considering flow with velocity field $\bar{v}(\bar r)=v \hat a_z$ \si{m.s^{-1}} and the degradation reaction \eqref{deg1}, the molecular diffusion can be described by partial differential equation (PDE) \cite{Grindrod}
\begin{align}\label{fick}
&D{\nabla ^2}C(\bar r,t|{{\bar r}_{\rm tx}},{t_0}) - \bar v(\bar r)\cdot {\bar \nabla C}(\bar r,t|{{\bar r}_{\rm tx}},{t_0})
 - {k_{d}}C(\bar r,t|{{\bar r}_{\rm tx}},{t_0}) + S(\bar r,t,{\bar{r}_{\rm tx}},t_0) \\
 &= \frac{{\partial C(\bar r,t|{{\bar r}_{\rm tx}},{t_0})}}{{\partial t}},\nonumber
\end{align}
where $C(\bar r,t|{{\bar r}_{\rm tx}},{t_0})$ denotes the concentration of molecules at point $\bar r$ and time $t$ given the impulsive point source $S(\bar r,t,{\bar{r}_{\rm tx}},t_0)=\frac{{\delta (\rho  - {\rho _{\rm tx}})}}{\rho }\delta(\varphi-\varphi_{\rm tx})\delta(z-z_{\rm tx}) \delta(t-t_0)$, $\nabla ^2$ and $\bar \nabla$ are Laplace and gradient operators, respectively. \color{black}
By employing Laplace and gradient operators in cylindrical coordinate system, \color{black} \eqref{fick} is rewritten as
\begin{align}\label{Eq}
&D\frac{{{\partial ^2}C(\bar r,t|{{\bar r}_{\rm tx}},{t_0})}}{{\partial {\rho ^2}}} + \frac{D}{\rho }\frac{{\partial C(\bar r,t|{{\bar r}_{\rm tx}},{t_0})}}{{\partial \rho }} + \frac{D}{{{\rho ^2}}}\frac{{{\partial ^2}C(\bar r,t|{{\bar r}_{\rm tx}},{t_0})}}{{\partial {\varphi ^2}}}\\\nonumber
& + D\frac{{{\partial ^2}C(\bar r,t|{{\bar r}_{\rm tx}},{t_0})}}{{\partial {z^2}}} - v\frac{{\partial C(\bar r,t|{{\bar r}_{\rm tx}},{t_0})}}{{\partial z}} - {k_{d}}C(\bar r,t|{{\bar r}_{\rm tx}},{t_0})\\ \nonumber
 &+ \frac{{\delta (\rho  - {\rho _{\rm tx}})}}{\rho }\delta (\varphi-\varphi_{\rm tx} )\delta (z-z_{\rm tx})\delta (t - {t_0}) = \frac{{\partial C(\bar r,t|{{\bar r}_{\rm tx}},{t_0})}}{{\partial t}}.\nonumber
\end{align}
The irreversible ligand-receptor reaction over the cylinder boundary given in \eqref{deg2} is characterized by the third type (Robin) boundary condition of \cite{Elka} \footnote{Since the condition is over the inner boundary, i.e.,  $C(\bar r,t|{{\bar r}_{\rm tx}},{t_0})$ is the concentration for $\rho \leq \rho_c$, the negative sign is appeared in right side.}
\begin{equation}\label{BD1}
D\frac{{\partial C(\bar r,t|{{\bar r}_{\rm tx}},{t_0})}}{{\partial \rho }}\mid_{\bar r=(\rho_c,z,\varphi)}=-k_{f} C(\rho_c,z,\varphi,t|{{\bar r}_{\rm tx}},{t_0}).
\end{equation}
Concentration function $C(\bar r,t|{{\bar r}_{\rm tx}},{t_0})$ that satisfy \eqref{Eq} subject to the boundary condition \eqref{BD1} is called concentration Green's function (CGF) of diffusion.


The CGF could be represented by the product of a one-dimensional and a two-dimensional Green's function as \cite{HG}
\begin{equation}\label{C}
  {C(\bar r,t|{{\bar r}_{\rm tx}},{t_0})}=C_{\rho\varphi}(\rho,\varphi ,t|\rho_{\rm tx},\varphi_{\rm tx},{t_0})C_z(z,t|z_{\rm tx},{t_0}),
\end{equation}
where $C_{\rho\varphi}(\rho,\varphi ,t|\rho_{\rm tx},\varphi_{\rm tx},{t_0})$ and  $C_z(z,t|z_{\rm tx},{t_0})$ are radial-azimuthal and axial direction Green's functions. Thereby, \eqref{Eq} reduces to the following two independent equations \cite{HG}:
\begin{align}\label{Cr}
&D\frac{{{\partial ^2}{C_{\rho \varphi }}(\rho ,\varphi ,t|{{\bar r}_{\rm tx}},{t_0})}}{{\partial {\rho ^2}}} + \frac{D}{\rho }\frac{{\partial {C_{\rho \varphi }}(\rho ,\varphi ,t|{{\bar r}_{\rm tx}},{t_0})}}{{\partial \rho }} + \frac{D}{{{\rho ^2}}}\frac{{{\partial ^2}{C_{\rho \varphi }}(\rho ,\varphi ,t|{{\bar r}_{\rm tx}},{t_0})}}{{\partial {\varphi ^2}}}\\
& + \frac{{\delta (\rho  - {\rho _{\rm tx}})}}{\rho }\delta (\varphi  - {\varphi _{\rm tx}})\delta (t - {t_0}) = \frac{{\partial {C_{\rho \varphi }}(\rho ,\varphi ,t|{{\bar r}_{\rm tx}},{t_0})}}{{\partial t}},\nonumber
\end{align}
\begin{align}\label{Cz}
&D\frac{{{\partial ^2}{C_z}(z,t|{z_{\rm tx}},{t_0})}}{{\partial {z^2}}} - v\frac{{\partial {C_z}(z,t|{z_{\rm tx}},{t_0})}}{{\partial z}} - {k_{d}}{C_z}(z,t|{z_{\rm tx}},{t_0}) + 
\delta (z-z_{\rm tx})\delta (t - {t_0}) \\
&= \frac{{\partial {C_z}(z,t|{z_{\rm tx}},{t_0})}}{{\partial t}}.\nonumber
\end{align}
Applying \eqref{C} in boundary condition \eqref{BD1} results in
\begin{equation}\label{CrB}
  D\frac{{\partial {C_{\rho\varphi }}(\rho,\varphi ,t|{\rho_{\rm tx}},\varphi_{\rm tx},{t_0})}}{{\partial \rho}}\mid_{\rho = \rho_c} = -k_{f} {C_{\rho\varphi }}(\rho_c,\varphi ,t|{\rho_{\rm tx}},\varphi_{\rm tx},{t_0}).
\end{equation}
Therefore,  $C_{\rho\varphi}(\rho,\varphi ,t|\rho_{\rm tx},\varphi_{\rm tx},{t_0})$  is the solution of PDE \eqref{Cr} subject to the boundary condition \eqref{CrB} and $C_z(z,t|z_{\rm tx},{t_0})$ is the solution of PDE \eqref{Cz}.
In the next two subsections, we present the solutions for these two equations.
\subsection{Derivation of radial-azimuthal CGF}
In this subsection, we solve \eqref{Cr} subject to the boundary condition given in \eqref{CrB}. The source term of $\frac{{\delta (\rho  - {\rho _{\rm tx}})}}{\rho }\delta (\varphi  - {\varphi _{\rm tx}})\delta (t - {t_0})$ in \eqref{Cr} is equivalent to considering initial condition of
\begin{equation} \label{sepin}
C_{\rho \varphi }(\rho ,\varphi ,t=t_0|{{\bar r}_{\rm tx}},{t_0})=\frac{{\delta (\rho  - {\rho _{\rm tx}})}}{\rho }\delta (\varphi  - {\varphi _{\rm tx}}).
\end{equation}
By considering this initial condition and removing the source term in \eqref{Cr}, a homogeneous PDE is obtained which can be solved by the technique of separation of variables {\cite{HG}}. By applying $C_{\rho \varphi }(\rho ,\varphi ,t|{{\bar r}_{\rm tx}},{t_0})=R(\rho|\rho_{\rm tx})\Phi(\varphi|\varphi_{\rm tx})T(t|t_0)$ in \eqref{Cr} without the source term and also in boundary condition \eqref{CrB}\color{black}, dividing both sides by $R(\rho|\rho_{\rm tx})\Phi(\varphi|\varphi_{\rm tx})T(t|t_0)$, and some simple manipulation, we obtain
\begin{equation}\label{sep}
   \frac{\rho^2R''(\rho|\rho_{\rm tx})}{R(\rho|\rho_{\rm tx})}+\frac{\rho R'(\rho|\rho_{\rm tx})}{R(\rho|\rho_{\rm tx})}-\frac{\rho^2 T'(t|t_0)}{DT(t|t0)}=\frac{\Phi''(\varphi|\varphi_{\rm tx})}{-\Phi(\varphi|\varphi_{\rm tx})}\overset{(a)}{=}\alpha,
\end{equation}
subject to the boundary condition
\begin{equation}\label{sep3B}
   DR'(\rho|\rho_{\rm tx})\mid_{\rho=\rho_c} = -{k_{f}}R(\rho_c|\rho_{\rm tx}),
 \end{equation}
where equality with constant $\alpha$ in (a) holds because of separation of variables in the left and right hand sides of the first equality, and prime ($'$) and double prime ($''$) symbols represent the first and second derivatives of the function with respect to its single variable \color{black}.
Thereby, we have the following ordinary differential equation
\begin{equation}\label{sep1}
\Phi''(\varphi|\varphi_{\rm tx})+\alpha \Phi(\varphi|\varphi_{\rm tx})=0.
\end{equation}
Considering that the concentration is periodic with period $2\pi$ in terms of $\varphi$ variable and it is a symmetric function respect to $\varphi=\varphi_{\rm tx}$, the acceptable values for $\alpha$ are $\alpha=n^2, \forall n\in \mathbb{Z}_{+}$ where $\mathbb{Z}_{+}$ denotes non-negative integer values. Correspondingly, $\Phi_n(\varphi|\varphi_{\rm tx})=G_n \cos(n(\varphi-\varphi_{\rm tx}))$ is a possible solution for \eqref{sep1} in which $G_n$ is unknown constant.

Considering $\alpha=n^2$ in \eqref{sep} and some simple manipulation, we obtain
\begin{equation}\label{sep2}
   \frac{DR_n''(\rho|\rho_{\rm tx})}{R_n(\rho|\rho_{\rm tx})}+\frac{D R_n'(\rho|\rho_{\rm tx})}{\rho R_n(\rho|\rho_{\rm tx})}-\frac{Dn^2}{\rho^2}=\frac{T'_n(t|t_0)}{T_n(t|t_0)}\overset{(b)}{=}-\gamma_n^2,
\end{equation}
where equality with constant in (b) holds because of separation of variables in the left and right hand sides of the first equality.
Note that only a negative constant on the right side is possible, since a nonnegative constant leads to unbounded function $T(t|t_0)$ and correspondingly unbounded concentration function of time which is impossible. Defining $\lambda_n=\gamma_n/\sqrt{D}$, \eqref{sep2} results in
 \begin{equation}\label{sep3}
  \rho^2 R_n''(\rho|\rho_{\rm tx})+\rho R_n'(\rho|\rho_{\rm tx})+ (\lambda _{n}^2\rho^2 - {n^2})R_n(\rho|\rho_{\rm tx})= 0,
 \end{equation}
subject to the boundary condition
\begin{equation}\label{sep3B2}
DR_n'(\rho|\rho_{\rm tx})\mid_{\rho=\rho_c} = -{k_{f}}R_n(\rho_c|\rho_{\rm tx}).
\end{equation}
Equation \eqref{sep3} is Bessel's equation with general solution \cite{HG}
\begin{equation}\label{Bessl}
R_n(\rho|\rho_{\rm tx})=A_nJ_n(\lambda_{n}\rho)+B_nY_n(\lambda_{n}\rho),
\end{equation}
where $J_n(.)$ and $Y_n(.)$ are $n^{th}$ order Bessel functions of first and second kind, respectively, for every positive value $\lambda_n$. Since $Y_n(\lambda_{n}\rho)$ is singular at $\rho=0$, we set $B_n=0$. Applying $R_n(\rho|\rho_{\rm tx})=A_nJ_n(\lambda_{n}\rho)$ in the boundary condition \eqref{sep3B2}, it is required to have 
\begin{equation}\label{lambda}
D{\lambda _{n}}{J_n}'({\lambda _{n}}\rho_c) =-k_{f}J_n({\lambda _{n}}\rho_c).
\end{equation}
Thereby, each root of \eqref{lambda} is an acceptable $\lambda_n$ value except $\lambda_n=0$ for $n>0$ which leads to the trivial solution of $R_n(\rho|\rho_{\rm tx})=0$. We note that $\lambda_0=0$ is a root for the boundary condition \eqref{lambda}, when $k_f=0$, which leads to the solution $R_0(\rho|\rho_{\rm tx})=A_0$. \color{black}
Denoting the $m^{th}$ root of the above equation by $\lambda_{nm}$, $R_{nm}(\rho_c|\rho_{\rm tx})=A_{nm}J_n(\lambda_{nm}\rho)$ is  a solution for \eqref{sep3} with boundary condition \eqref{sep3B2}. Given $\lambda_{nm}$ and considering the implicit condition of  $\lim_{t\to \infty} T(t|t_0)=0$, $T(t|t_0)=I_{nm}e^{-D\lambda_{nm}^2 (t-t_0)}u(t-t_0)$ satisfies \eqref{sep2}, where $I_{nm}$ is an unknown constant and $u(.)$ is the step function. Therefore, we have
\begin{align}\label{Crphif}
{C_{\rho \varphi }}(\rho ,\varphi ,t|{{\bar r}_{\rm tx}},{t_0}) = \sum\limits_{n=0}^\infty \sum\limits_{m = 1}^\infty
{H_{nm}J_n(\lambda_{nm}\rho)\cos (n(\varphi-\varphi_{\rm tx}))e^{-D\lambda_{nm}^2 (t-t_0)}u(t-t_0)},
\end{align}
 where $H_{nm}=G_nA_{nm}I_{nm}$ which is unknown and should be determined by applying the initial condition given in \eqref{sepin}. Delta functions $\delta (\varphi  - {\varphi _{\rm tx}})$ and $\frac{{\delta (\rho  - {\rho _{\rm tx}})}}{\rho }$ can be expanded as following, respectively, \cite{GF}
 \begin{equation}\label{dphi}
 \delta (\varphi  - {\varphi _{\rm tx}}) =\sum\limits_{n=0}^\infty  {L_n \cos (n(\varphi  - \varphi _{\rm tx}))},
\end{equation}
where $L_0=\frac{1}{2\pi}$ and $L_n=\frac{1}{\pi}, n\geq 1$, and
 \begin{equation}\label{drho}
 \frac{{\delta (\rho  - {\rho _{\rm tx}})}}{\rho } = \sum\limits_{m = 1}^\infty  {\frac{{{J_n}({\lambda _{nm}}{\rho _{\rm tx}})}}{{{N_{nm}}}}} {J_n}({\lambda _{nm}}\rho ),
 \end{equation}
 in which
\begin{equation}\label{Nmn}
N_{nm}^{} = \int_{0}^{\rho_c} {\rho J_n^2} ({\lambda _{nm}}\rho)d\rho
 = \frac{{{{\rho_c}^2}}}{2}(J_n^2({\lambda _{nm}}\rho_c) - J_{n - 1}^{}({\lambda _{nm}}\rho_c)J_{n + 1}^{}({\lambda _{nm}}\rho_c)).
\end{equation}

By applying \eqref{Crphif}-\eqref{drho} to initial condition \eqref{sepin} and comparing left and right sides of the equation, we obtain
 \begin{equation}
 H_{nm}={\frac{{{J_n}({\lambda _{nm}}{\rho _{\rm tx}})}}{{{N_{nm}}}}}L_n, \;\;n\geq0, m\geq 1.
  \end{equation}

\subsection{Derivation of axial CGF}
To solve \eqref{Cz}, we take Fourier transform of both sides of \eqref{Cz} in terms of axial variable $z$ and we obtain
\begin{equation}\label{Cz0}
(- D{\beta ^2} - {k_{d}} - j\beta v)\tilde C_z(\beta,t\mid z_{\rm tx},{t_0}) + \delta (t - {t_0})= \\
\frac{{\partial \tilde C_z(\beta,t\mid z_{\rm tx},{t_0})}}{{\partial t}},
\end{equation}
 where $\tilde C_z(\beta,t\mid z_{\rm tx},{t_0})$ denotes the Fourier transform of  $C_z(z,t\mid z_{\rm tx},{t_0})$, i.e., we have
\begin{equation}\label{CzFou}
{C_z}(z,t\mid{z_{\rm tx}},{t_0}) =\frac{1}{{2\pi }}\int_{ - \infty }^\infty  {\tilde C_z(\beta,t\mid z_{\rm tx},{t_0})} {e^{j\beta z}}d\beta.
\end{equation}
  Given $\beta$, this is an ordinary differential equation in terms of $t$ which can be easily solved as
 \begin{equation}\label{Czs}
  \tilde C_z(\beta,t\mid z_{\rm tx},{t_0}) = \frac{e^{( - D{\beta ^2} - {k_{d}} - j\beta v)(t - {t_0})}}{2\pi}.
\end{equation}

By taking the inverse Fourier transform of \eqref{Czs}, we obtain
\begin{equation}\label{Cz00}
{C_z}(z,t|{z_{\rm tx}},{t_0}) = \\
\frac{1}{{{\sqrt{(4\pi D(t - {t_0}))}}}}{e^{\frac{{ - {{(z - {z_{\rm tx}} - v(t - {t_0}))}^2}}}{{4D(t - {t_0})}} - {k_d}(t - {t_0})}} u(t-t_0).
\end{equation}
Substituting \eqref{Crphif} and \eqref{Cz00} in \eqref{C}, the CGF of diffusion in cylinder is obtained as follows
\begin{align}\label{Cf}
C(\bar r,t|{{\bar r}_{\rm tx}},{t_0}) &= \frac{1}{{{\sqrt{(4\pi D(t - {t_0}))}}}}{e^{\frac{{ - {{(z - {z_{\rm tx}} - v(t - {t_0}))}^2}}}{{4D(t - {t_0})}} - {k_d}(t - {t_0})}}\\
 &\times \sum\limits_{n=0}^\infty \sum\limits_{m = 1}^\infty
{{\frac{{L_n{J_n}({\lambda _{nm}}{\rho _{\rm tx}})}}{{{N_{nm}}}}}J_n(\lambda_{nm}\rho)\cos (n(\varphi-\varphi_{\rm tx}))e^{-D\lambda_{nm}^2 (t-t_0)}u(t-t_0)},\nonumber
\end{align}
where $L_0=\frac{1}{2\pi}$ and $L_n=\frac{1}{\pi}, n\geq 1$.

We have remarked two properties of the obtained CGF in below. 
\begin{remark}
The CGF \eqref{Cf} for the fluid without flow ($v=0$) reveals reciprocity property. In other words, by interchanging the location of observation point $(\rho,z,\varphi)$ and point source transmitter $(\rho_{\rm tx},z_{\rm tx},\varphi_{\rm tx})$, the CGF does not change, when $v=0$. The reciprocity property of CGF results in the reciprocity of the corresponding channel of DMC which may be exploited in analyzing and designing DMC networks. 
\end{remark}
\begin{remark}
Assuming transmitter located on the cylinder axis ($\rho_{\rm tx}=0$) leads to the azimuthal symmetry ( $\frac{{{\partial ^2}C}}{{\partial {\varphi ^2}}}=0$). Thereby, we have only zero order Bessel function in the series of \eqref{Cf}, i.e., CGF is reduced to 
\begin{align}\label{AS}
C(\bar r,t|{{\bar r}_{tx}},{t_0}) &= \frac{1}{{{\sqrt{(4\pi D(t - {t_0}))}}}}{e^{\frac{{ - {{(z - {z_{\rm tx}} - v(t - {t_0}))}^2}}}{{4D(t - {t_0})}} - {k_d}(t - {t_0})}}\\
 &\times \sum\limits_{m = 1}^\infty  {\frac{{{J_0}({\lambda _{0m}}{\rho _{tx}}){J_0}({\lambda _{0m}}\rho )}}{{{N_{0m}}}}} {e^{ - \gamma _{0m}^2(t - {t_0})}} u(t-t_0).\nonumber
\end{align}
\end{remark}

\section{Characterization of DMC Channel}

According to the analysis in previous section, assuming an impulsive point source of molecule release $\frac{{\delta (\rho  - {\rho _{\rm tx}})}}{\rho }\delta(\varphi-\varphi_{\rm tx})\delta(z-z_{\rm tx})$, the CGF, $C(\bar{r},t|\bar{r}_{\rm tx},t_0)$, is given by \eqref{Cf}. %
Therefore, given an arbitrary transmitter (not necessarily point source or with instantaneous release) of $S(\bar r,t), \bar r \in \Omega$, the concentration at arbitrary observation point $\bar{r}=(\rho,\varphi,z)$ is obtained as follows
\begin{equation}\label{out}
\iiint_\Omega \int_{-\infty}^{+\infty} C(\bar r,t|\bar r',t') S(\bar r',t') dt' \rho' d\rho' dz' d\varphi',
\end{equation}
where $C(\bar r,t|\bar r',t')$ is given by \eqref{Cf}.

We note that the differential equation in \eqref{Eq} that introduces the system with point source input at $(\rho_{\rm tx},\varphi_{\rm tx},z_{\rm tx})$ and CGF output, $C(\bar{r},t|\bar{r}_{\rm tx},t_0)$ is linear time invariant. Therefore, for the special case of point source transmitter located at $\bar r_{\rm tx}$ with molecule release rate of $s(t)$, $S(\bar r',t')=s(t')\frac{{\delta (\rho'  - {\rho _{\rm tx}})}}{\rho }\delta(\varphi'-\varphi_{\rm tx})\delta(z'-z_{\rm tx})$, and $C(\bar{r},t|\bar{r}'_{\rm tx},t')$ given by \eqref{Cf}, \eqref{out} reduces to $s(t)*C(\bar{r},t|\bar{r}_{\rm tx},t_0=0)$ where $*$ is convolution operator with respect to $t$.\color{black}

%

\subsection{Statistics of receiver signal}
Assume a point source transmitter located at $\bar{r}_{\rm tx}=(\rho_{\rm tx},\varphi_{\rm tx},z_{\rm tx})$ and a transparent receiver where the set of points inside the receiver is denoted by $\Omega_{\rm rx}$. \color{black}   
To analyze the noise in the received signal, we first provide the probability density function (pdf) of the observation time of an individual molecule at the receiver in the following. Releasing one molecule at time $t=0$ from the point transmitter at $\bar{r}_{\rm tx}=(\rho_{\rm tx},\varphi_{\rm tx},z_{\rm tx})$ is equivalent to the impulsive point source, $\frac{1}{\rho} \delta(\rho-\rho_{\rm tx})\delta(\varphi-\varphi_{\rm tx})\delta(z-z_{\rm tx})\delta(t)$. \color{black}  
Hence, ${C(\bar {r},t|{{\bar {r}}_{\rm tx}},{t_0=0})}$ given in \eqref{Cf} can be interpreted as the probability density of presence of molecule at point $\bar{r}$ and time $t$.
Therefore, the pdf of observation of the molecule in a transparent receiver at time $t$ is obtained as
\begin{align}\label{eqlemm2}
p_{\rm obs}(t)=\iiint_{\Omega_{\rm rx}}{{C(\bar {r},t|{{\bar {r}}_{\rm tx}},{t_0=0})} \rho d\rho \varphi dz},
\end{align}
where $\Omega_{\rm rx}$ denotes the set of points inside the receiver.

The release rate of a realistic transmitter is stochastic, since chemical reactions involve in the release of molecules that are inherently stochastic \cite{Arjmandi2013}. In fact, transmitter can control the average release rate of molecules. For instance, for ion channel and ion pump biosynthetic modulators proposed in \cite{Ion}, \cite{Pump}, the release rate of molecules has been modeled as a Poisson process $\textbf{s}(t)\sim \mathrm{Poisson} (s(t))$ where $s(t)$ is the average modulated signal. \footnote{We note that the bold font is used to denote the random variables or processes, in this paper.\color{black}} 

Assume the transmitter intends to transmit average modulated signal $s(t)$  for $t\in [0, T]$ corresponding with an intended input symbol. Correspondingly, the release rate of molecules is modeled as Poisson process
\begin{align}\label{TXmodel}
\textbf{s}(t)\sim \mathrm{Poisson} (s(t)).
\end{align}
%

It is proved in \cite{Ion}, the number of the molecules observed at the receiver at time $t\in[0,T]$, $\textbf{y}(t)$, originating from the molecules released in interval $[0,T]$ follows a Poisson distribution with mean
\begin{align}
y(t)&=\int_{0}^{T}{s(\tau)p_{\rm obs}(t-\tau) d\tau}=s(t)*p_{\rm obs}(t)\\
&=\iiint_{\Omega_{\rm rx}}{{s(t)*C(\bar {r},t|{{\bar {r}}_{\rm tx}},{t_0=0})} \rho d\rho \varphi dz},\nonumber
\end{align}
where the last equality is resulted by substituting $p_{\rm obs}(t)$ from \eqref{eqlemm2}.   \color{black}Assuming an spherical receiver with very small radius $R_{\rm rx}$ compared to distance between transmitter and receiver, the concentration variations inside the receiver is negligible and probability density of observation time given in \eqref{eqlemm2} can be approximated by $\frac{4\pi}{3}R_{\rm rx}^3 {C(\bar r_{\rm rx},t|{\bar r_{\rm tx}},{t_0})}$, where $\bar r_{\rm rx}$ is the center of the receiver and $R_{\rm rx}$ is the receiver volume \cite{Mahfuz14}. Therefore, we can approximate
\begin{align}
y(t)=\frac{4\pi}{3}R_{\rm rx}^3(s(t)* {C(\bar r_{\rm rx},t|{\bar r_{\rm tx}},{t_0})}).
\end{align}

\subsection{Intersymbol interference (ISI)}
We characterized the received signal at the receiver due to transmitted signal in the current time slot, in the last subsection. Now, we explain how the residual ISI from the previous time slots can be incorporated in the receiver output.

Let $i$ denote the time slot number such that $i=0$ refers to the current time slot $[0,T]$ and $i>0$ denoted a previous time slot $[-iT,-(i-1)T]$. Assume the average modulated signal in time slot $i$ corresponding with the input symbol for transmission  in this time slot is denoted by $s_i(t+iT)$. The number of molecules observed in the current time slot, $i=0$, at time $t\in[0,T]$, originating from a previously transmitted signal $s_i(t+iT)$ in time slot $i>0$, $[-iT,(1-i)T]$, is denoted by $\textbf{I}_{i}(t)$. Similar to the derivation of the stochastic of ${\textbf{y}}(t)$ , it is straightforward to show that $\textbf{I}_{i}(t)$ is Poisson distributed with mean \cite{Ion}
\begin{align}
I_i(t)&=\int_{-iT}^{(1-i)T }s_i(\tau+iT)p_{\rm obs}(t-\tau)d\tau\\
&=\int_{0}^{T}s_i(\tau)p_{\rm obs}(iT+t-\tau)d\tau\nonumber\\
&=s_i(iT+t)*p_{\rm obs}(iT+t).\nonumber
\end{align}

Assume the diffusion channel has memory of length $M$ time slots. The total ISI affecting the receiver output originating from $M$ previously transmitted symbols in the current time slot, $\textbf{I}$, is given by
\begin{align}\label{ISI}
\textbf{I}=\sum_{i=1}^{M}\textbf{I}_i,
\end{align}
which follows a Poisson distribution since the $\textbf{I}_i$ are mutually independent Poisson RVs for $i\in \{1,\cdots,M\}$.
Therefore, given the current transmitted modulated signal, $\textbf{s}_0(t)$, the receiver output in the current time slot is $\textbf{y}_R=\textbf{y}+\textbf{I}$ which is a Poisson distributed RV with mean
\begin{align}
{y}_R&={s_0(t)*p_{\rm obs}(t)}+\sum_{i=1}^{M}{s_i(iT+t)*p_{\rm obs}(iT+t)}=\sum_{i=0}^{M}{s_i(iT+t)*p_{\rm obs}(iT+t)}.  
\end{align}

\subsection{Performance analysis for on-off keying modulation}
In order to evaluate the DMC system performance based on proposed analysis, a simple on-off keying modulation is adopted where bits 1 and 0 are represented by the average modulated signals $N\delta(t)$ and $0$, respectively. In other words, assuming transmission of bit 1, the transmitter release molecules instantaneously at the beginning of the time slot where the number of released molecules is a Poisson RV with mean $N$. The transparent receiver counts the number of molecules inside the receiver volume at sampling time $t_s$ (which maximize $p_{\rm obs}(t)$) in each time slot.
The receiver uses the observed sample to decide about the transmitted bit.

Given the transmitted bits $B_i={b}_i, i\in\{0,1,\cdots,M\}$, the average modulated signal in time slot $i\in\{0,1,\cdots,M\}$ is $s_i(t+iT)=Nb_i\delta(t+iT)$. As shown in the last subsection, $\textbf{y}_R$ is a Poisson distributed RV, i.e.,\color{black}  
\begin{align}
&{\rm Pr}(\textbf{y}_R=y| b_0, {b}_1,\cdots,{b}_M)=\frac{e^{-\mathbb{E}(\textbf{y}_R| b_0, {b}_1,\cdots,{b}_M)}(\mathbb{E}(\textbf{y}_R| b_0, {b}_1,\cdots,{b}_M))^y}{y!},
\end{align}
in which
\begin{align}
&\mathbb{E}(\textbf{y}_R| b_0, {b}_1,\cdots,{b}_M)=\sum_{i=0}^{M}{{b}_iN\delta(iT+t)*p_{\rm obs}(iT+t)}=\sum_{i=0}^{M}{{b}_iNp_{\rm obs}(iT+t)},
\end{align}
where ${\rm Pr}(\cdot)$ and $\mathbb{E}(\cdot)$ denote probability function and expectation operator, respectively.\color{black} 

For the error probability analysis, we assume a genie-aided decision feedback (DF) detector \cite{Mosayebi2014}, where a genie informs the detector of the previously transmitted bits, i.e., $\hat{B}_i=B_i,\;i=1,\cdots,M$. Assuming the decoder knows the correct values of the previously transmitted bits, i.e., $B_i={b}_i, i\in\{1,\cdots,M\}$, and ${\rm Pr}(B_0=1)={\rm Pr}(B_0=0)=\frac{1}{2}$, the Maxiumum-A-Posteriori (MAP) detector for bit $B_0$ given receiving $Y=y$ molecules in the current time slot becomes
\begin{align}\label{map}
\hat{B}_0=\mathrm{arg}\max_{b_0\in\{0,1\}}{{\rm Pr}(\textbf{y}_R=y| b_0, {b}_1,\cdots,{b}_M)},
\end{align}
where $\hat{B}_0$ denotes the estimated transmitted bit in the current time slot. Since, the previously transmitted bits $B_i=b_i,\; i\in\{1,\cdots,M\}$, are not known in practice, previous decisions $\hat{B}_i=\hat{b}_i,\; i\in\{1,\cdots,M\}$, have to be used in \eqref{map} instead.

Simplifying \eqref{map} leads to a threshold decision rule based on the receiver output in the current time slot, $y$, \cite{Ion} i.e., $\hat{B}_0=0$, if $y\leq   {\rm Thr}$, and $\hat{B}_0=1$, if $y> {\rm Thr}$, where
\begin{align}\label{map_thr}
\mathrm{Thr}=\frac{{Np_{\rm obs}(t_s)}}{\ln \left(1+\frac{{Np_{\rm obs}(t_s)}}{\sum_{i=1}^{M}{Nb_ip_{\rm obs}(iT+t_s)}}\right)}.
\end{align}
The error probability of this detector is given by
\begin{align}\label{errorprob}
P_{\rm error}=\left(\frac{1}{2}\right)^{M+1}\sum_{b_0,\cdots ,b_M} {{\rm Pr}(E|b_0,b_1,\cdots,b_M)}, 
\end{align}
where $E$ is an error event, and we have 
\begin{align}\label{cond_error}
&{\rm Pr}(E|b_0,b_1,\cdots,b_M)=\nonumber\\
&\sum_{y \underset{b_0=0}{\overset{b_0=1}{\lessgtr}} {\rm Thr}}{\frac{e^{-\mathbb{E}(\textbf{y}_R| b_0,b_1,\cdots,b_M)}(\mathbb{E}(\textbf{y}_R| b_0, b_1,\cdots,b_M))^y}{y!}}.
 \end{align}
 \color{black}

\section{Simulation and Numerical Results}
In this section, we investigate the effect of system parameters on CGF for diffusion in the considered biological cylindrical environment. Moreover, the performance of the point-to-point DMC system over this channel is evaluated.
To confirm the proposed analysis of the CGF, we employ a particle based simulator (PBS).
In the PBS, the molecule locations are known and the molecules move independently in the 3-dimensional space. In each dimension (Cartesian coordinates), the displacement of a molecule in $\Delta t$ s is modeled as a Gaussian RV with zero mean and variance $2D\Delta t$. Subject to the degradation reaction given in \eqref{deg1}, a molecule may be degraded (removed) from the environment during a time step $\Delta t$ s with probability of $k_{d}\Delta t$ \cite{Elka}.
If a molecule hits the boundary which is covered by receptor proteins characterized by \eqref{deg2}, the molecule may bind with receptor $R$ and produce molecule $AR$ with probability $k_{f} \sqrt{\frac{\pi \Delta t}{D}}$ and may be reflected with probability of $1-k_f \sqrt{\frac{\pi \Delta t}{D}}$ \cite{Elka}. We note that employing this probability for simulating the boundary condition results in quantitatively accurate PBS, when the simulation time steps or adsorption coefficients are very small (more precisely $k_f \sqrt{\frac{\Delta t}{2D}}\ll 1/ \sqrt{2\pi}$)\cite{Andrew10}. \color{black}
\begin{table}
	\begin{center}
		 		 \caption{ِDMC system parameters used in simulations} \label{table1}
		\begin{tabular}{|c|c|c|}
			\hline
			\bfseries{Parameter} & \bfseries{Variable} & \bfseries{Value}   \\
			\hline \hline
			Diffusion coefficient & $D$ & $ 10^{-9}$ \si{m^2.s^{-1}}\\
			\hline
			Cylinder radius & $\rho_c$ & $5, 7.5, 10, 15$ \si{\mu m}\\
			\hline
			Point source transmitter location & $(\rho_{\rm tx},z_{\rm tx},\varphi_{\rm tx})$ & $(3\si{\mu m},0,0)$\\
			\hline
			Degradation reaction constant inside   & $k_d$ & $0, 20$ \si{s^{-1}}\\the cylinder & &  \\
			\hline
			Ligand-receptor reaction constant over & $k_f$ & $0, 100, \infty$ \si{\mu m.s^{-1}}\\ the surface & & \\
			\hline
			Flow velocity & $v$ & 65 \si{\mu m.s^{-1}}\\
			\hline
			Receiver radius & $R_{\rm rx}$ & $0.5\; \si{\mu m}$ \\
			\hline
			Number of transmitted molecules for bit '1'& $N$ & $5 \times 10^4$ \\
			\hline
			Time step in PBS& $\Delta t$ & $10^{-5} $\si{s} \\
			\hline
			Number of realizations in PBS & Not represented & $10^{7} $ \\
			\hline
			Employed eigenvalues ($\lambda_{nm}$) to obtain CGF & $n$ and $m$ & $n\leq 3$ and $m\leq 5$\color{black} \\
			\hline
		\end{tabular}
	\end{center}
\end{table}
The point source transmitter is located at $(\rho_{\rm tx},z_{\rm tx},\varphi_{\rm tx})=(3\mu m,0,0)$ and Diffusion coefficient is $D=10^{-9}$ \si{m^2.s^{-1}}. Although the CGF given in \eqref{Cf} is an infinite series, but it practically converges to the PBS result by considering a limited number of eigenvalues ($\lambda_{mn}$), which depends on the system parameters. For the system parameters used in this paper, $n\leq 3$ and $m\leq 5$ lead to enough accurate CGF compared to the PBS.
\color{black} The system parameters adopted for the analytical and simulation results are given in Table \ref{table1}.

\begin{figure}
\center
\includegraphics[width=15 cm,height=9 cm]{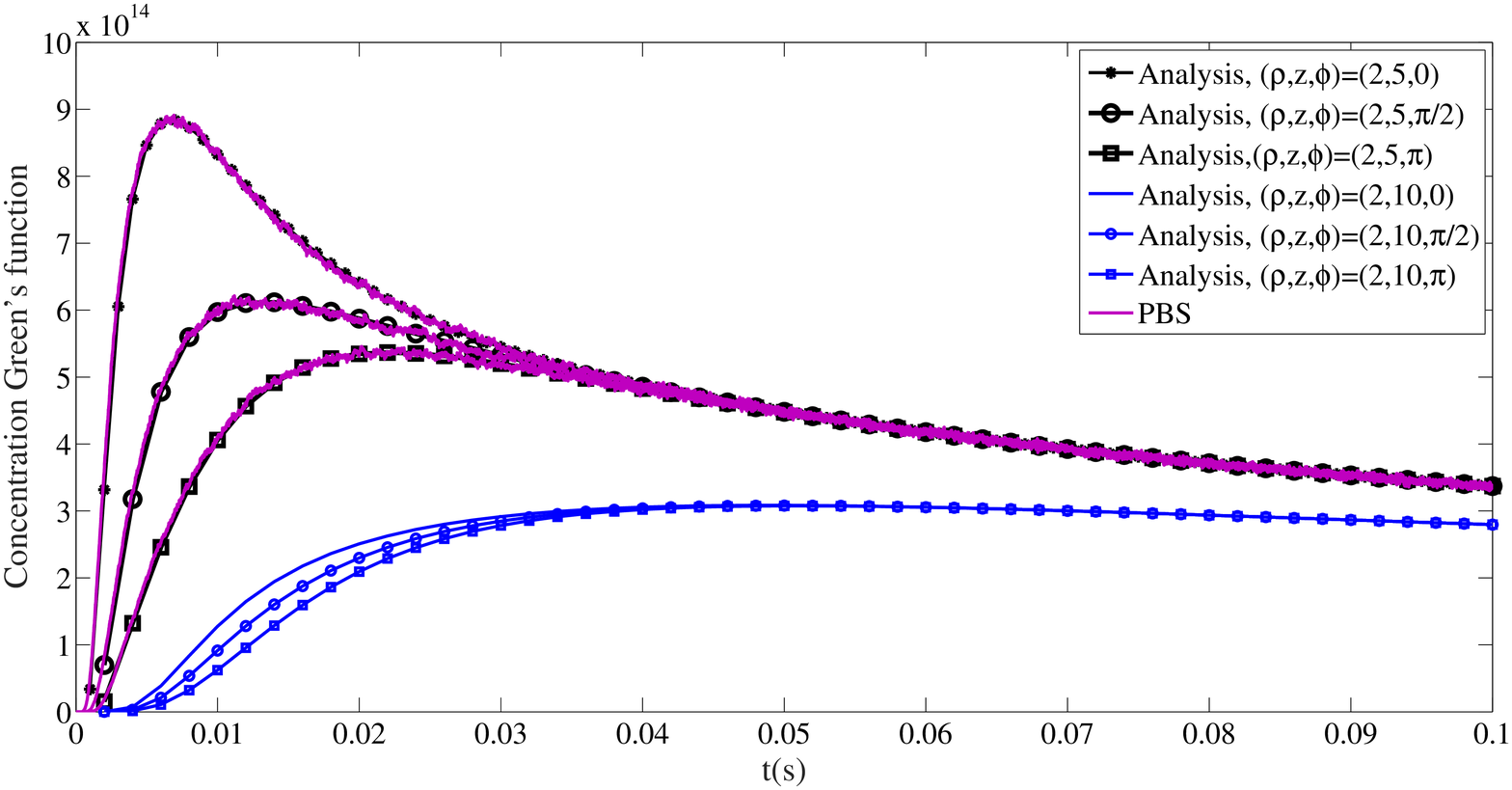}
	\setlength{\abovecaptionskip}{-0.5 cm}
 \caption{CGF obtained from analysis and PBS for observation point with $\rho=2\mu m$ and different axial and azimulthal coordinates, $z=5,10 \mu m$ and $\varphi=0,\pi/2,\pi$, when point transmitter is located at $(3\mu m,0,0)$, $k_f=0$, $v=0$, and $k_{d}=0$.\color{black}}
\label{Fig1}
\end{figure}

Fig. \ref{Fig1} compares the CGF obtained from our analysis given in \eqref{Cf} and PBS for observation points located at $\rho=2$ \si{\mu m}, $z=5$ \si{\mu m} with different azimuthal coordinates $\varphi=0,\pi/2,$ and $\pi$ when $\rho_c=5$ \si{\mu m}. To examine only the effect of azimuthal coordinate in our analysis, we have considered reflecting boundary ($k_f=0$), $v=0$, and $k_{d}=0$. It is observed that the PBS confirms the proposed analysis which captures concentration variations in azimuthal coordinate in addition to radial and axial coordinates. Also, Fig. \ref{Fig1} depicts the CGF obtained from analysis for observation points with $\rho=2$ \si{\mu m}, $z=10$ \si{\mu m} for different azimuthal coordinates of $\varphi=0,\pi/2,$ and $\pi$. Comparing with CGF curves for observation points at $\rho=2$ \si{\mu m}, $z=5$ \si{\mu m}, it is deduced that CGF variation in azimuthal coordinate decreases by increasing axial distance between point source and observation point ($|z-z_{\rm tx}|$). This occurs because the effect of the axial coordinate on diffusion is dominant for large values of $|z-z_{\rm tx}|$, as \eqref{Cf} shows.

\begin{figure}
\center
\includegraphics[width=15 cm,height=9 cm]{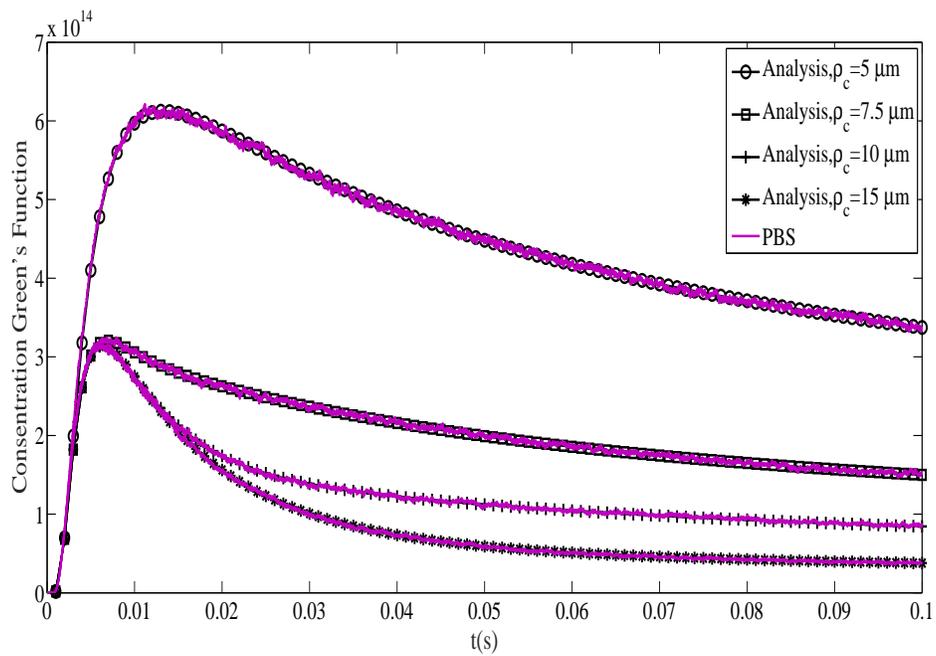}
	\setlength{\abovecaptionskip}{-0.5 cm}
 \caption{CGF obtained from analysis and PBS for different cylinder radius, $\rho_c=5,7.5,10,15$ $\mu m$, when point transmitter and observation point are located at $(3\mu m,0,0)$ and $(2\si{\mu m},5\si{\mu m},\pi/2)$, respectively, $k_f=0$, $v=0$, and $k_{d}=0$.\color{black}}
\label{Fig2}
\end{figure}

Fig. \ref{Fig2} depicts the CGF obtained from the analysis in \eqref{Cf} and the PBS for different cylinder radius values of $\rho_c=5,10,$ and $15$ \si{\mu m} when the observation point is located at $(2\si{\mu m},5\si{\mu m},\pi/2)$, $v=0$, $k_{d}=0$, $k_{f}=0$.  It is observed that the PBS confirms the proposed analysis. Fig. \ref{Fig2} show that CGF is significantly amplified by decreasing the radius of the cylinder. Thereby, a cylinder environment with smaller radius result in stronger received signals at the receiver and consequently may improve the DMC system performance.

\begin{figure}
\center
\includegraphics[width=15 cm,height=9 cm]{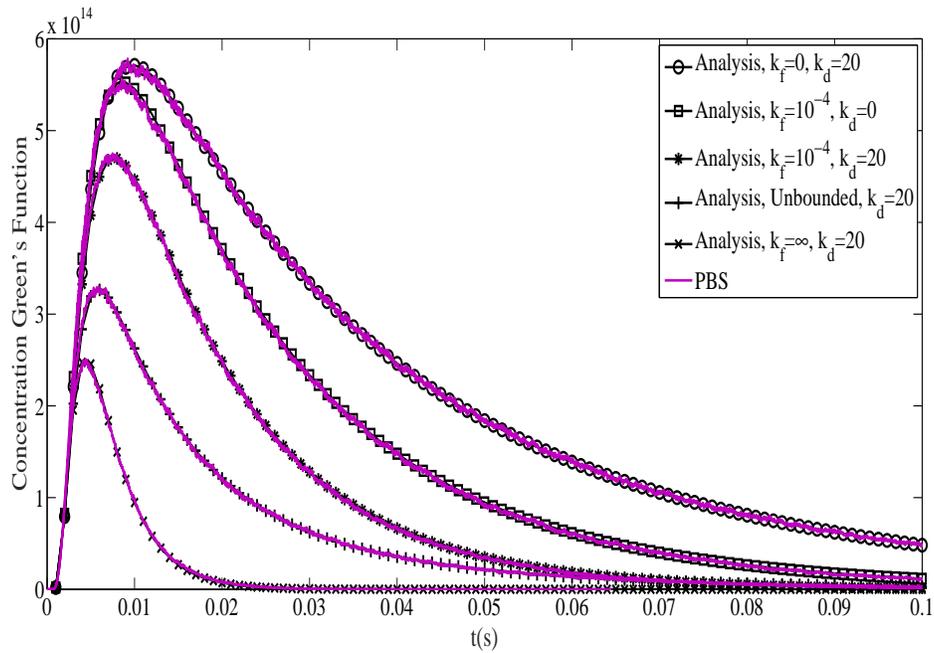}
	\setlength{\abovecaptionskip}{-0.5 cm}
 \caption{CGF obtained from analysis and PBS for diffusion in unbounded and cylindrical environment for different $k_{d}$ and $k_{f}$ values, when point transmitter and observation point are located at $(3\mu m,0,0)$ and $(2\si{\mu m},5\si{\mu m},\pi/2)$, respectively, $\rho_c=5$ \si{\mu m}, and $v=65$ \si{\mu m.s^{-1}}.\color{black}}
\label{Fig4}
\end{figure}

In Fig. \ref{Fig4}, the CGF in the presence of degradation reaction ($k_{d}=0,20$), with different boundary conditions including absorbing $(k_{f}\to \infty$), reflective ($k_{f}=0$), partially absorbing ($k_f=10^{-4}$), and unbounded ($\rho_c\to \infty$) boundaries is examined. The CGF obtained from \eqref{Cf} and PBS has been depicted for observation point located at $(2\si{\mu m},5 \si{\mu m},\pi/2)$ when $\rho_c=5$ \si{\mu m}, $v=65$ \si{\mu m.s^{-1}}. In all scenarios PBS confirms the proposed analytic results. It is observed that degradation inside the environment and binding with the receptors over the surface weakens the CGF (correspondingly the gain of the diffusion channel) from one side and shorten the tail of CGF (correspondingly the memory of diffusion channel) from the other side. Therefore, a trade-off between the gain and memory of the diffusion channel arises in the presence of degradation and partially absorbing boundary. As expected, CGF in unbounded scenario has higher amplitudes compared to the absorbing boundary, since the molecules hitting to the boundary are removed and do not have the chance to return to the environment. On the other hand, it has lower amplitude compared to the reflective boundary, since the molecules movement are limited within the boundary for reflective boundary which leads to higher concentration inside the cylinder. Similarly, the trade-off between the channel gain and memory is deduced by comparing the CGFs for absorbing and partially absorbing boundaries and also the CGFs for reflective and partially absorbing boundaries. \color{black}

\begin{figure}
\center
\includegraphics[width=15 cm,height=9 cm]{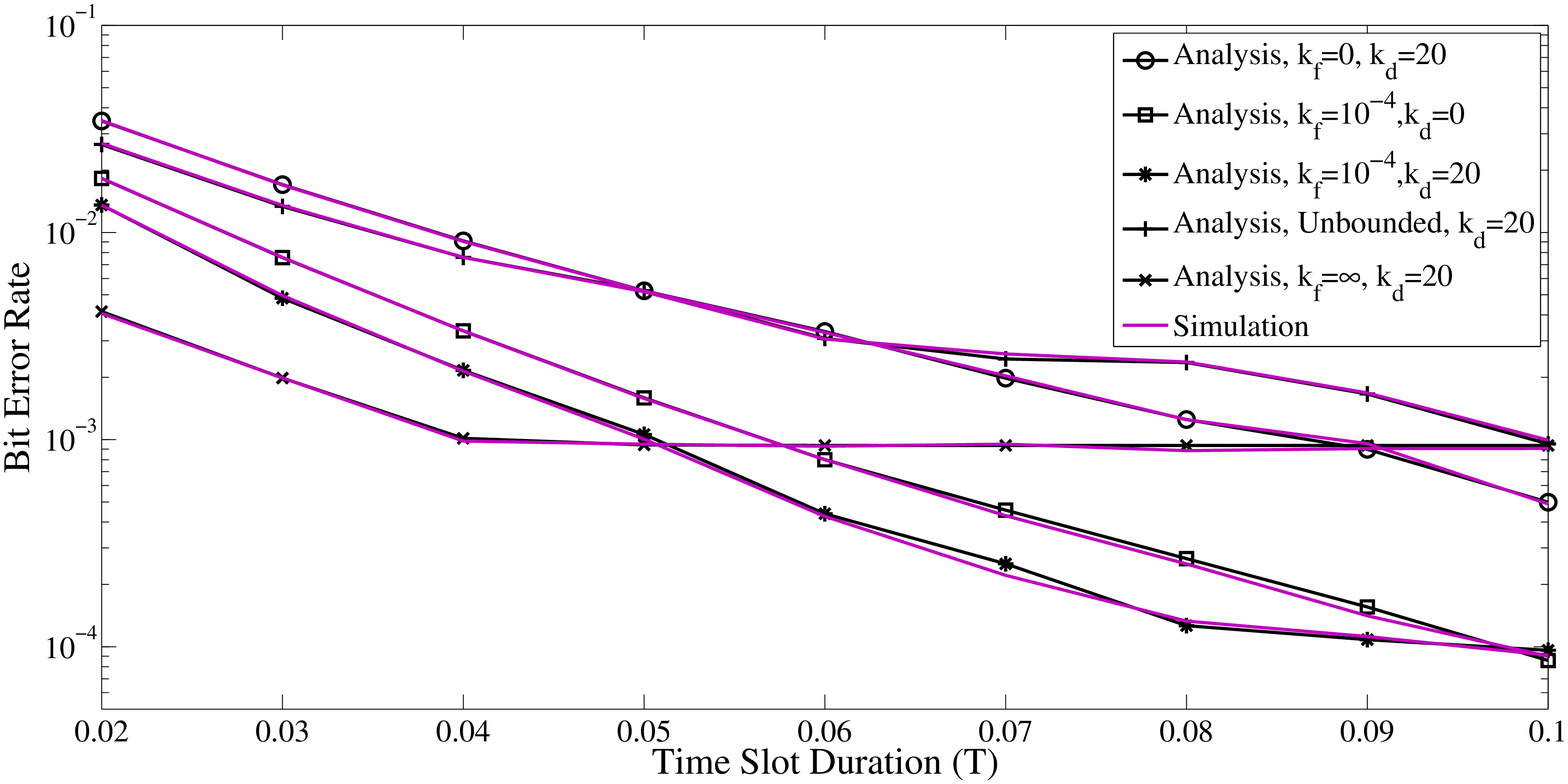}
	\setlength{\abovecaptionskip}{-0.5 cm}
 \caption{BER of the DMC system corresponding with scenarios in Fig. \ref{Fig4}. }
\label{Fig5}
\end{figure}

Corresponding with different scenarios used in Fig. \ref{Fig4}, the performance of DMC system in terms of error probability has been shown in Fig. \ref{Fig5} where observation probability in each scenario is given by $P_{\rm obs}=\frac{4\pi}{3}R_{\rm rx}^3 {C(\bar r_{\rm rx},t|{\bar r_{\rm tx}},{t_0})}$. A simple on-off keying modulation scheme is considered where 0 and 1 are represented by releasing 0 and $N=5\times 10^4$ molecules (on average) by the transmitter, respectively. The center of the transparent spherical receiver with radius $R_{\rm rx}=0.5$ \si{\mu m} is located at $(2\si{\mu m},5 \si{\mu m},\pi/2)$. The receiver observes the number of molecules inside its volume at sampling time in which observation probability is maximized. The receiver uses this observation value to decide about received signal. The error probability of different scenarios obtained from \eqref{errorprob} has been depicted versus time slot duration, $T$, being verified by Monte Carlo simulation with $10^6$ bits. 

In all scenarios, the BER is a decreasing function of time slot duration, since for a shorter time slot duration (higher transmission rate), a higher memory and ISI is encountered.
It is also observed that absorbing boundary with $k_{d}=20$ results in a higher and lower BER for $T<0.09$ and $T>0.09$ compared to reflective boundary with $k_{d}=20$, respectively. It reveals the trade off between the gain and memory of the diffusion channel resulted from absorbing boundary mentioned above. In fact, the effect of channel memory is dominant for smaller $T$ values; thus absorbing boundary results smaller BER than reflective boundary. Also, the effect of channel gain is dominant for higher $T$ values; thus the reflective boundary case improves BER. In addition, we observe that partially absorbing boundary leads to a higher BER for $T<0.05$ compared to the absorbing boundary when $k_{d}=20$ which implies the effect of channel memory is dominant. On the other hand, partially absorbing boundary outperforms BER compared to the absorbing boundary for $T>0.05$ when $k_{d}=20$ which reveals the effect of the channel gain is dominant. Similarly, we observe that the scenario with degradation reaction ($k_d=20$) outperforms the BER compared to the scenario in the absence of degradation reaction ($k_d=0$) for $T<0.095$ which reveals the effect of channel memory is dominant for $T<0.095$ in the trade-off between the channel gain and channel memory. \color{black}      

\begin{figure}
\center
\includegraphics[width=15 cm,height=9 cm]{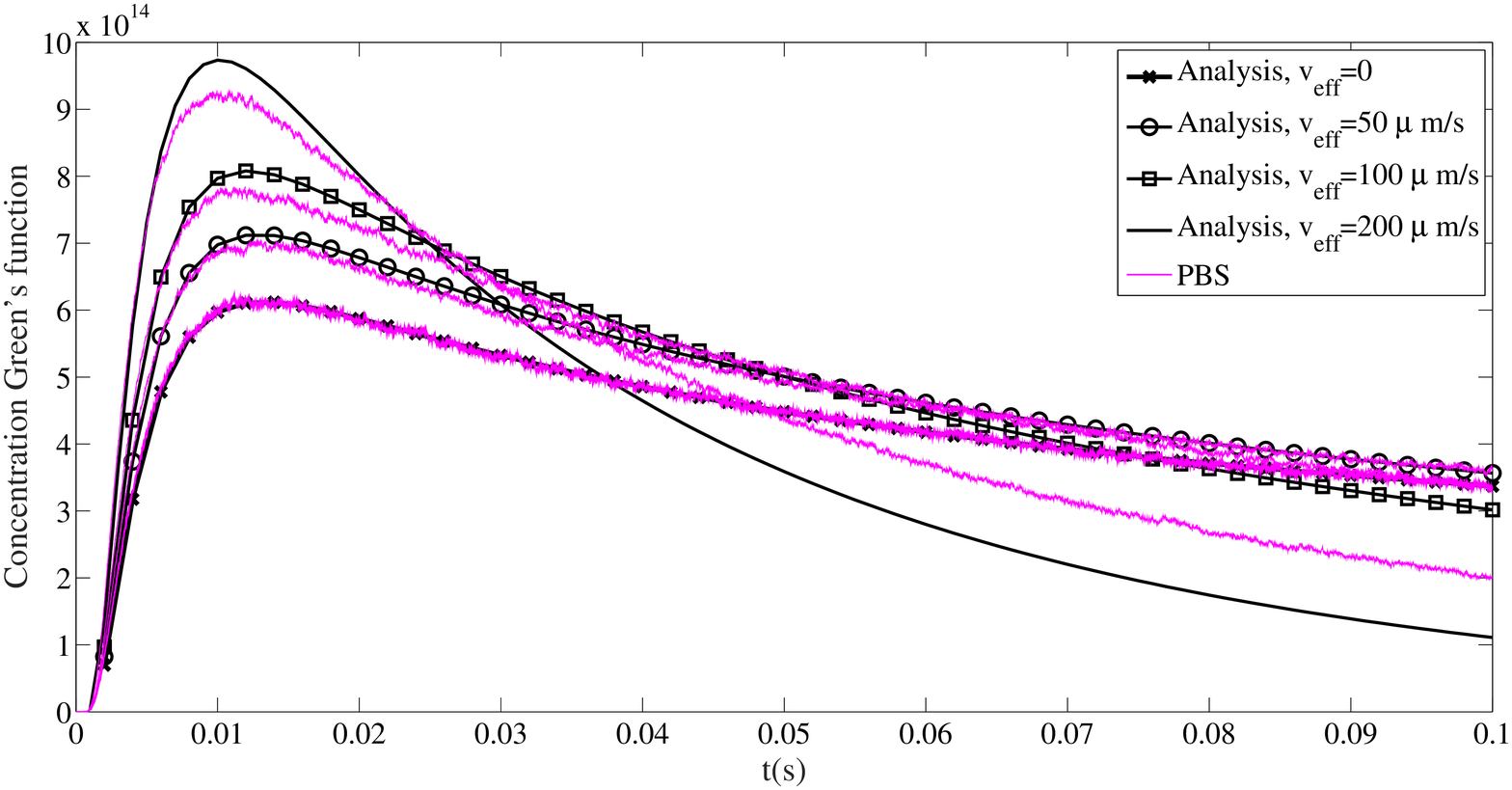}
	\setlength{\abovecaptionskip}{-0.5 cm}
 \caption{CGF obtained from analysis and PBS for different flow effective velocities $v_{\rm eff}=0, 50,100$ and 200 $\mu ms^{-1}$, when point transmitter and observation point are located at $(3\mu m,0,0)$ and $(2\si{\mu m},5\si{\mu m},\pi/2)$, respectively, $\rho_c=5$ \si{\mu m}, $k_f=0$, and $k_{d}=0$ \color{black}.}
\label{Fig3}
\end{figure}

In our analysis for CGF, we assumed a constant velocity flow which is not a realistic assumption. Poiseuille model is a well-known flow model for cylinder in which $\bar v(\bar r)=2v_{\rm eff}(1-\frac{\rho^2}{\rho_c^2}) \hat a_z$ \si{m.s^{-1}}.
Fig. \ref{Fig3} compares the CGF obtained from our analysis given in \eqref{Cf}, with PBS with Poiseuille flow model.
To have a fair comparison, we have considered the constant velocity $v$ in our analysis equals to the average velocity in Poiseuille model, i.e., $v=\int_{0}^{\rho_c} {2v_{\rm eff}(1-\frac{\rho^2}{\rho_c^2})d\rho}=\frac{4}{3}v_{\rm eff}$. We have considered different effective velocities $v_{\rm eff}=0, 50,100$ and 200 \si{\mu m.s^{-1}}, when the observation point is located at $(2\si{\mu m},5 \si{\mu m},\pi/2)$, $\rho_c=5$ \si{\mu m}, $k_{f}=0$, $k_{d}=0$.
It is observed that the proposed analysis coincides the PBS for zero velocity and well approaches PBS results for enough small effective velocity values. More accurately, given the cylinder radius $\rho_c$, the distribution of velocity in cylinder diameter has less variation (deviation from mean velocity) for smaller $v_{\rm eff}$ values which leads to better approximation of PBS results compared to higher $v_{\rm eff}$ values. Similarly, we can argue that given $v_{\rm eff}$, increasing the radius of the cylinder results less variation (average deviation from mean velocity) leading to better approximation of PBS results. \color{black}

\section{Conclusion}
Inspired from the blood vessel structures in the body, a biological cylindrical environment was considered for DMC. Considering degradation inside the cylinder, irreversible receptor proteins over the boundary, and uniform flow, the concentration Green's function of diffusion in this environment was analytically derived which takes into account asymmetry in all radial, axial and azimuthal coordinates. Correspondingly, information channel between the DMC transmitter and receiver was characterized. 
The effect of different system parameters on the channel response was examined. It was deduced that channel response variation in azimuthal coordinate decreases by increasing axial distance between point source and observation point. Also, it was observed that a cylinder environment with smaller radius results in stronger received signals at the receiver and consequently may improve the DMC system performance. Moreover, a trade-off between the gain and memory of the diffusion channel in the presence of degradation and partially absorbing boundary was revealed. 
\color{black}
To evaluate the DMC system in the biological cylinder, a simple on-off keying modulation scheme was adopted and corresponding error probability was derived. We observed that degradation mechanism and partially absorbing boundary can be utilized to mitigate ISI resulted by the previous time slots. The proposed model can be used to design, optimize, and evaluate the DMC inside the blood vessels for biomedical applications. Obtaining CGF in biological cylinder with reversible receptor proteins over its boundary, reactive receiver and non-unifrom flow is left for future works\color{black}.

\newpage


\begin{thebibliography}{00}
\bibitem{Akyldiz2011}
I. F. Akyildiz, F. Brunetti, and C. Blázquez, ``Nanonetworks: A new communication paradigm,'' \textit{Computer Networks}, vol. 52, no. 12, pp. 2260-2279, Aug. 2008.

\bibitem{Pierobon2011}
M.~Pierobon and I.~Akyildiz, ``A physical end-to-end model for molecular communications in nanonetwork,'' \textit{IEEE Journal on Selected Areas in Communications}, vol. 28, no. 4, pp. 602-611, May 2010.

\bibitem{Nakano12}
T. Nakano, M. J. Moore, F.Wei, A. V. Vasilakos, and J. Shuai, ``Molecular communication and networking: Opportunities and challenges,'' \textit{IEEE Transactions on NanoBioscience}, vol. 11, no. 2, pp. 135-148, Jun. 2012.
%
\bibitem{Chahibi2013}
Y. Chahibi, M. Pierobon, S. O. Song, and I. F. Akyildiz, ``A molecular communication system
model for particulate drug delivery systems,''  \textit{IEEE Transactions on Biomedical Engineering}, vol. 60, no. 12,
pp. 3468-3483, Dec. 2013.
\bibitem{Nakano13}
T. Nakano, A. W. Eckford, and T. Haraguchi, \textit{Molecular Communication}, Cambridge, U.K.: Cambridge University Press, 2013.
\bibitem{Farsad16}
N. Farsad, H. B. Yilmaz, A. Eckford, C. B. Chae, and W. Guo, ''A comprehensive
survey of recent advancements in molecular communication," \textit{IEEE Communications Surveys and Tutorials}, vol. 18, no. 3, pp. 1887-1919, Feb. 2016.


\bibitem{Kuran10}
M. S. Kuran, H. B. Yilmaz, T. Tugcu, and B. Özerman, ``Energy model for communication via diffusion in nanonetworks,'' \textit{Nano Communication Networks}, vol. 1, no. 2, pp. 86-95, Jun. 2010.


\bibitem{Noise2011}
M. Pierobon and I. F. Akyildiz,``Diffusion-based noise analysis for molecular communication in nanonetworks,'' \textit{IEEE Transactions on Signal Processing}, vol. 59, no. 6, pp. 2532-2547, Jun. 2011.

\bibitem{Pier14}
M. Pierobon and I. F. Akyildiz, ``A statistical physical model of interference
in diffusion-based molecular nanonetworks,'' \textit{IEEE Transactions on Communications}, vol. 62, no. 6, pp. 2085-2095, Jun. 2014.
\bibitem{Bazar14}
M. H. Bazargani and D. Arifler, ``Deterministic model for pulse amplifcation in diffusion-based molecular communication,'' \textit{IEEE Communications Letters},
vol. 18, no. 11, pp. 1891-1894, Nov. 2014.
\bibitem{Noel14}
A. Noel, K. Cheung, and R. Schober, ``Optimal receiver design for diffusive molecular communication with flow and additive noise,'' \textit{IEEE Transactions
on NanoBioscience}, vol. 13, no. 3, pp. 350-362, Sep. 2014.

\bibitem{Aijaz15}
A. Aijaz and A. H. Aghvami, ``Error performance of diffusion-based molecular communication using pulse-based modulation,'' \textit{IEEE Transactions
on NanoBioscience}, vol. 14, no. 1, pp. 146-151, Jan. 2015.

\bibitem{Kilinc13}
D. Kilinc and O. B. Akan, ``Receiver design for molecular communication,'' \textit{IEEE Journal of Selected Areas in Communications}, vol. 31, no. 12, pp. 705-714, Dec. 2013.
\bibitem{Wang15}
X. Wang, M. D. Higgins, and M. S. Leeson, ``Relay analysis in molecular communications with time-dependent concentration,'' \textit{IEEE Communications Letters}, vol. 19, no. 11, pp. 1977-1980, Nov. 2015.
\bibitem{Mahfuz14}
M. U. Mahfuz, D. Makrakis, and H. T. Mouftah, ``A comprehensive study of sampling-based optimum signal detection in concentration-encoded
molecular communication,'' \textit{IEEE Transactions
on NanoBioscience}, vol. 13, no. 3, pp. 208-222, Sep. 2014.

\bibitem{Assaf}
S. S. Assaf, S. Salehi, R. G. Cid-Fuentes, J. Solé-Pareta, and E. Alarcón, ``Influence of neighboring absorbing receivers upon the inter-symbol interference in a diffusion-based molecular communication system," \textit{Nano Communication Networks}, vol. 14, pp.40-47, 2017.
\color{black}
\bibitem{Alzubi18}
M. M. Al-Zu’bi, M. M., A. S. Mohan, ''Modeling of ligand-receptor protein interaction in biodegradable spherical bounded biological micro-environments," \textit{IEEE Access}, vol. 6, 25007-25018, May 2018.


\bibitem{Dinc182}
F. Dinc, B. C. Akdeniz, A. E. Pusane, T. Tugcu, ''Impulse response of the channel with a spherical absorbing receiver and a spherical reflecting boundary. arXiv preprint arXiv:1804.03383, 2018.


\bibitem{Four}
 R. L. Fournier, \textit{Basic transport phenomena in biomedical engineering},
CRC press, 2017. \color{black}

\bibitem{Farsad12}
N. Farsad, A. W. Eckford, S. Hiyama, Y. Moritani, ''On-chip molecular communication: Analysis and design," \textit{IEEE Transactions on NanoBioscience}, vol. 11, no. 3, pp. 304-314, Sep. 2012.

\bibitem{Kuran13}
M. S. Kuran, H. B. Yilmaz, and T. Tugcu, ''A tunnel-based approach for signal shaping in molecular communication" \textit{IEEE International Conference on Communications Workshops}, pp. 776-781, Jun. 2013.

\bibitem{Feli13}
L. Felicetti, M. Femminella, and G. Reali, ''Establishing digital molecular communications in blood vessels," \textit{First IEEE International Black Sea Conference on Communications and Networking (BlackSeaCom)}, pp. 54-58, Jul. 2013.

\bibitem{Feli14}
L. Felicetti, M. Femminella, G. Reali, P. Gresele, M. Malvestiti, J.N. Daigle, “Modeling CD40-based molecular communications in blood vessels,” IEEE Transactions on Nanobioscience, Vol. 13, no. 3, 2014.
\color{black}
\bibitem{Turan18}
M. Turan, M. S. Kuran, H. B. Yilmaz, H. B., I. Demirkol, T. Tugcu, ''Channel model of molecular communication via diffusion in a vessel-like environment considering a partially covering receiver. arXiv preprint arXiv:1802.01180, 2018.





\bibitem{Felicetti13}
L. Felicetti, M. Femminella, G. Reali, ``Simulation of molecular signaling in blood vessels: Software design and application to atherogenesis," \textit{Nano Communication Networks}, Vol. 4, no. 3, 2013.

\bibitem{Unterweger}
H. Unterweger, J. Kirchner, W. Wicke, A. Ahmadzadeh, D. Ahmed, V. Jamali, C. Alexiou, G. Fischer, and R. Schober, ``Experimental molecular communication testbed based on magnetic nanoparticles in duct flow," \textit{IEEE SPAWC}, 2018. 
\bibitem{Schefer}
M. Sch{\"a}fer, W. Wicke, R. Rabenstein, and R. Schober, ``Analytical models for particle diffusion and flow in a horizontal cylinder with a vertical force" \textit{ arXiv preprint arXiv:1803.10848}, 2018.

\color{black}
\bibitem{Wayan17}
W. Wicke, T. Schwering, A. Ahmadzadeh, V. Jamali, A. Noel, and R. Schober, ''Modeling duct flow for molecular communication." arXiv preprint arXiv:1711.01479, 2017.



\bibitem{Dinc18}
F. Dinc, F., B. C. Akdeniz, A. E. Pusane, T. Tugcu, ''A general analytical solution to impulse response of 3-D microfluidic channels in molecular communication," arXiv preprint arXiv:1804.10071, 2018.

\bibitem{Cliff}
W.J. Cliff, \textit{Blood vessels,} No. 6. CUP Archive, 1976.

\bibitem{Adam14}
A. Noel, K. C. Cheung, and R. Schober, "Improving receiver performance of diffusive molecular communication with enzymes,'' \textit{IEEE Transactions on NanoBioscience}, vol. 13, no 1, pp. 31-43, Mar. 2014.

\bibitem{Nelson}
P. Nelson, \textit{Biological Physics: Energy, Information, Life}, 1st ed. San
Francisco, CA, USA: Freeman, 2008.


\bibitem{GF}
D. G. Duffy, \textit{Green's function with applications}. CRC Press, 2015.


\bibitem{syncArj}
L. Felicetti, M. Femminella, G. Reali, T. Nakano, and A. V. Vasilakos, ''TCP-like molecular communications,'' \textit{IEEE Journal on Selected Areas in Communications}, vol. 32, no. 12, pp. 2354-2367, Dec. 2014.



%
\bibitem{Grindrod}
P. Grindrod, The theory and applications of reaction-diffusion equations:
patterns and waves. Clarendon Press, 1996.

\bibitem{Elka}
Y. Deng, A. Noel, M. Elkashlan, A. Nallanathan, and K. C. Cheung, ''Modeling and simulation of molecular communication systems with a reversible adsorption receiver" \textit{IEEE Transactions on Molecular, Biological and Multi-Scale Communications}, vol. 1, no. 4 , pp.347-362, Dec. 2015.

\bibitem{HG}
K. D. Cole, J. V. Beck, A. Haji-Sheikh, and B. Litkouhi, \textit{Heat conduction using Green’s functions}. CRC Press, 2010.
\bibitem{Arjmandi2013}
H. Arjmandi, A. Gohari, M. Nasiri-Kenari, and Farshid Bateni, ``Diffusion based nanonetworking: A new modulation technique and performance analysis,'' \textit{IEEE Communications Letters}, vol. 17, no. 4, pp. 645-648, Apr. 2013.

\bibitem{Ion}
H. Arjmandi, A Ahmadzadeh, R. Schober, and M. N.  Kenari, ``Ion channel based bio-synthetic modulator for diffusive molecular communication,'' \textit{IEEE Transactions on Nanobioscience}, vol. 15, no. 5, pp. 418-432, Jul. 2016.
\bibitem{Pump}
H. Arjmandi, A Ahmadzadeh, R. Schober, and M. N.  Kenari, ``Ion Pump based bio-synthetic modulator for diffusive molecular communication,'' \textit{SPAWC}, 2016.








%
%
%


\bibitem{Mosayebi2014}
R. Mosayebi, H. Arjmandi, A. Gohari, M. Nasiri Kenari, and U. Mitra, ``Receivers for diffusion-based molecular communication: Exploiting memory and sampling rate,'' \textit{IEEE Journal of Selected Areas in Communications,} vol. 32, no. 12, pp. 2368-2380, Dec. 2014.







\bibitem{Andrew10}
S. S. Andrews, ``Accurate particle-based simulation of adsorption, desorption and partial transmission." \textit{Physical biology} vol. 6, no. 4, Nov. 2009.
















%




%
%
%
%
%
%
%
%
%
%
%
%
%
%
%
%
%
%
%
%

\end{thebibliography}
\end{document}